\documentclass[final]{svjour3}
\usepackage{gensymb}
\usepackage{graphicx}
\usepackage{rotating}
\usepackage{amssymb}
\usepackage{mathptmx}
\usepackage{comment}
\usepackage{floatrow}
\usepackage{bm}
\usepackage[numbers]{natbib}
\makeatletter
\journalname{Journal of Low Temperature Physics}
\bibpunct{}{}{,}{s}{}{,}

\begin{document}

\newcommand{\hdblarrow}{H\makebox[0.9ex][l]{$\downdownarrows$}-}
\title{Optical Design and Characterization of 40-GHz Detector and Module for the BICEP Array}

\author{A.~Soliman$^1$ \and P.~A.~R.~Ade$^2$ \and Z.~Ahmed$^{3,4}$ \and M.~Amiri$^5$ \and D.~Barkats$^6$ \and R.~Basu Thakur$^1$ \and C.~A.~Bischoff$^7$ \and J.~J.~Bock$^{1,8}$ \and H.~Boenish$^6$ \and E.~Bullock$^9$ \and V.~Buza$^6$ \and J.~Cheshire$^9$ \and J.~Connors$^6$ \and J.~Cornelison$^6$ \and M.~Crumrine$^9$ \and A.~Cukierman$^4$ \and M.~Dierickx$^6$ \and L.~Duband$^{10}$ \and S.~Fatigoni$^5$ \and J.~P.~Filippini$^{11,12}$ \and G.~Hall$^9$ \and M.~Halpern$^5$ \and S.~Harrison$^6$ \and S.~Henderson$^{3,4}$ \and S.~R.~Hildebrandt$^8$ \and G.~C.~Hilton$^{13}$ \and H.~Hui$^1$ \and K.~D.~Irwin$^{3,4,13}$ \and J.~Kang$^4$ \and K.~S.~Karkare$^{6,14}$ \and E.~Karpel$^4$ \and S.~Kefeli$^1$ \and J.~M.~Kovac$^6$ \and C.~L.~Kuo$^{3,4}$ \and K.~Lau$^9$ \and K.~G.~Megerian$^8$ \and L.~Moncelsi$^1$ \and T.~Namikawa$^{15}$ \and H.~T.~Nguyen$^8$ \and R.~O'Brient$^{8,1}$ \and S.~Palladino$^7$ \and T.~Prouve$^{10}$ \and N.~Precup$^9$ \and C.~Pryke$^9$ \and B.~Racine$^6$ \and C.~D.~Reintsema$^{13}$ \and S.~Richter$^6$ \and B.~Schmitt$^6$ \and R.~Schwarz$^6$ \and C.~D.~Sheehy$^{16}$ \and A.~Schillaci$^1$ \and T.~St.~Germaine$^6$ \and B.~Steinbach$^1$ \and R.~V.~Sudiwala$^2$ \and K.~L.~Thompson$^{3,4}$  \and C.~Tucker$^2$ \and A.~D.~Turner$^8$ \and C.~Umilt\`{a}$^7$ \and A.~G.~Vieregg$^{14}$ \and A.~Wandui$^1$ \and A.~C.~Weber$^8$ \and D.~V.~Wiebe$^5$ \and J.~Willmert$^9$ \and W.~L.~K.~Wu$^{14}$ \and E.~Yang$^4$ \and K.~W.~Yoon$^4$ \and E.~Young$^4$ \and C.~Yu$^4$ \and C.~Zhang$^1$}

\institute{$^1$Department of Physics, California Institute of Technology, Pasadena, California 91125, USA
\\$^2$School of Physics and Astronomy, Cardiff University, Cardiff, CF24 3AA, United Kingdom
\\$^3$Kavli Institute for Particle Astrophysics and Cosmology, SLAC National Accelerator Laboratory, 2575 Sand Hill Rd, Menlo Park, California 94025, USA
\\$^4$Department of Physics, Stanford University, Stanford, California 94305, USA
\\$^5$Department of Physics and Astronomy, University of British Columbia, Vancouver, British Columbia, V6T 1Z1, Canada
\\$^6$Harvard-Smithsonian Center for Astrophysics, Cambridge, Massachusetts 02138, USA
\\$^7$Department of Physics, University of Cincinnati, Cincinnati, Ohio 45221, USA
\\$^8$Jet Propulsion Laboratory, Pasadena, California 91109, USA
\\$^9$Minnesota Institute for Astrophysics, University of Minnesota, Minneapolis, 55455, USA
\\$^{10}$Service des Basses Temp\'{e}ratures, Commissariat \`{a} lEnergie Atomique, 38054 Grenoble, France
\\$^{11}$Department of Physics, University of Illinois at Urbana-Champaign, Urbana, Illinois 61801, USA
\\$^{12}$Department of Astronomy, University of Illinois at Urbana-Champaign, Urbana, Illinois 61801, USA
\\$^{13}$National Institute of Standards and Technology, Boulder, Colorado 80305, USA
\\$^{14}$Kavli Institute for Cosmological Physics, University of Chicago, Chicago, IL 60637, USA
\\$^{15}$Leung Center for Cosmology and Particle Astrophysics, National Taiwan University, Taipei 10617, Taiwan
\\$^{16}$Physics Department, Brookhaven National Laboratory, Upton, NY 11973
\\ \email{asoliman@caltech.edu}}

\maketitle

\begin{abstract}
Families of cosmic inflation models predict a primordial gravitational-wave background that imprints B-mode polarization pattern in the Cosmic Microwave Background (CMB). High sensitivity instruments with wide frequency coverage and well-controlled systematic errors are needed to constrain the faint B-mode amplitude. We have developed antenna-coupled Transition Edge Sensor (TES) arrays for high-sensitivity polarized CMB observations over a wide range of millimeter-wave bands. BICEP Array, the latest phase of the BICEP/Keck experiment series, is a multi-receiver experiment designed to search for inflationary B-mode polarization to a precision $\sigma$(\textit{$r$}) between 0.002 and 0.004 after 3 full years of observations, depending on foreground complexity and the degree of lensing removal.
We describe the electromagnetic design and measured performance of BICEP Array's low-frequency 40-GHz detector, their packaging in focal plane modules, and optical characterization including efficiency and beam matching between polarization pairs.  We summarize the design and simulated optical performance, including an approach to improve the optical efficiency due to mismatch losses. We report the measured beam maps for a new broad-band corrugation design to minimize beam differential ellipticity between polarization pairs caused by interactions with the module housing frame, which helps minimize polarized beam mismatch that converts CMB temperature to polarization ($T \rightarrow P$) anisotropy in CMB maps. 


\end{abstract}

\section{Introduction}
Cosmic Microwave Background (CMB) radiation provides us with key information about the early universe from the epoch of recombination. Some models of inflation, which solve several fundamental issues in cosmology, predict a B-mode CMB polarization pattern at an amplitude within experimental reach. The amplitude of inflationary B-mode polarization is parametrized by the tensor-to-scalar ratio ($r$). Results from the BICEP/Keck experiment [1] currently constrain inflationary gravitational waves at $r < 0.06$ at 95\% confidence. BICEP Array (BA) is the latest phase in the BICEP/Keck project, providing high-sensitivity measurements over a range of frequencies from 30 GHz to 270 GHz. After 3 full years of observations, BICEP Array will measure primordial gravitational waves to a precision $\sigma$(\textit{$r$}) between 0.002 and 0.004, depending on foreground complexity and the degree of lensing removal. The first low frequency BICEP array receiver [2] will cover two bands at 30 and 40 GHz to measure and subtract Galactic synchrotron emission. We will deploy this receiver to South Pole by end of 2019.

This 30/40 GHz BICEP Array receiver [3] will have six 30 GHz modules and six 40 GHz modules in a checker board arrangement over its focal plane. Each detector module contains a 0.625 mm detector wafer, a $\lambda$/4 thick quartz anti-reflection tile over the silicon, and a $\lambda$/4 Niobium (Nb) back short [4]. Each pixel in the array couples incident optical radiation to a Transition Edge Sensor (TES) bolometer via microstrip lines in a dual-polarization 8X8 array of slot antennas. Each microstrip feed contains a third-order band-defining microstrip filter, and thermalizes electromagnetic power in a termination resistor integrated onto a released bolometer island with a TES sensor. Superconducting Quantum Interference Device (SQUID) amplifiers using Time Domain Multiplexing (TDM) read out the signal current through the voltage biased TES. 

     The devices measure a polarized CMB signal by taking the difference between a polarization pair of detectors (Pol A- Pol B) sharing a common dual-polarization antenna. Differential pointing between pairs can cause systematic  temperature to polarization ($T \rightarrow P$) leakage, generating a false polarization signal from temperature anisotropy. BICEP uses a deprojection algorithm to measure and remove the lowest-order leakage effects.  However, higher-order mismatch is present in past arrays at 95 GHz [5], dominated by electromagnetic interactions between the antenna and the module walls.  We aim to minimize this effect on the BICEP Array target sensitivity [2].  This interaction is especially problematic at 30 and 40 GHz, where the wavelengths are longest compared with the physical dimensions of the module, and where the highest fraction of pixels are located on the housing frame.
      
      This paper starts with the design and simulated optical performance of the detector at 40 GHz with an emphasis on improving the optical efficiency due to mismatch losses in the feed network. We also present the antenna far-field measurements of a new wide-band corrugated design [6] for a 40 GHz BICEP Array module. 

\section{Detector Antenna Design}

High sensitivity detector arrays with tight control of systematic effects are necessary for measuring the faint B-mode signal. We have developed arrays of 8X8 dual-polarized antenna-coupled detectors to achieve this goal. The vertical polarization (Pol A) and horizontal polarization (Pol B) antenna arrays allow an independent measurement of each CMB polarization.  The electromagnetic waves from each polarization coherently sum in the microstrip feed-network and pass through a three pole Chebyshev band pass filter to avoid atmospheric lines and to reject out-of-band radiation as shown in Fig. 1.  

To aid our antenna design work, we simulate the optical efficiency and the radiation pattern. The optical efficiency is maximized by minimizing the mismatch losses. The input impedance of the antenna is calculated by HFSS commercial software as shown in Fig. 2. The feed location on the slot antenna has been chosen to provide low and fabricable matching impedance. We minimized the return loss by matching the antenna radiation resistance with the microstrip impedance (via line width) and cancel the antenna reactance with a shunt capacitor.  This results in an averaged optical efficiency of 84\% over 25\% bandwidth [4].

\begin{figure}[htbp]
\begin{center}
\includegraphics[width=0.8\linewidth, keepaspectratio]{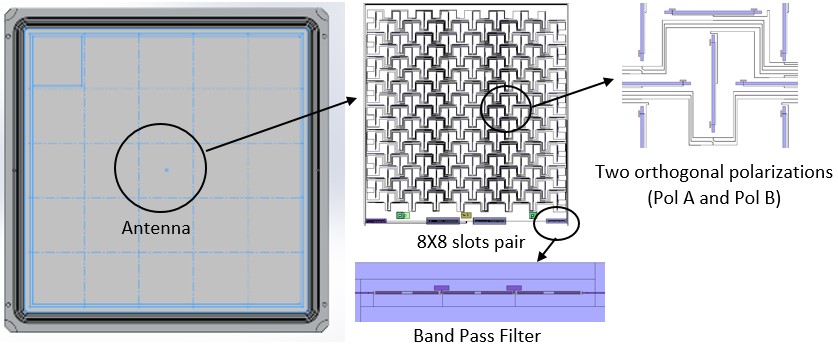}
\caption{ View of the 40 GHz module layout in a BICEP Array checker-board pattern. The 40 GHz module contains 5X5 antennas, each antenna has two orthogonal arrays of 8X8 slot pairs for dual polarization observations. Each microstrip feed contains three pole band-defining filter to define the upper and lower frequency cutoff of the science bands. (Color figure online.)}

\end{center}
\label{fig1}
\end{figure}

\begin{figure}[htbp]
\begin{center}
\includegraphics[width=0.9\linewidth, keepaspectratio]{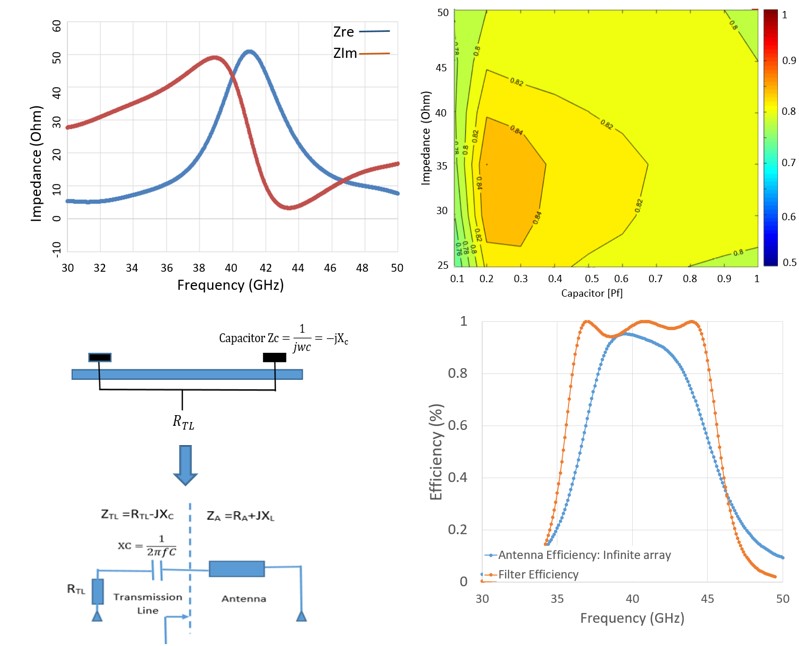}
\caption{{\it Top/Left:} The simulated real and imaginary impedance at the optimized feed point on the slot vs. frequency. {\it Top/Right:}The contour matching of the input microstrip impedance/capacitor to the antenna impedance. {\it Bottom/Left:}The equivalent matching circuit to the antenna. We used a series capacitor that shunt to ground to cancel out the inductive reactance impedance of the antenna.  {\it Bottom/Right:} The expected antenna/filter efficiency versus frequency. The measured end to end Optical  Efficiency  (OE) is  approximately between 30\% to 36\%. (Color figure online.)}

\end{center}
\label{fig2}
\end{figure}

\section{ Module Design and Performance}

The detector module mounts the detector array, anti-reflection wafer, Nb back short and readout to an aluminum frame and provides a compact structure that can fill BICEP Array cameras' large focal planes. However, the electromagnetic interaction between the frame and the antennas located at the edge of the metal frame can cause unwanted beam distortions.  For narrow band applications, quarter wavelength corrugations are typically used to suppress these interactions because the electrical short in the back of the corrugation transforms into an electromagnetic open at the front surface nearest the antenna.  Our novel broadband corrugation [6] uses seven corrugation slots of quarter-wavelength depths, alternating between $\lambda/4$ = 1.88 mm at 40 GHz and $\lambda$ = 2.5 mm at 30 GHz, with a total frame height of $H$ = 7.5 mm.  Fig. 3 shows the simulation of a 40 GHz beam in the presence of this broadband corrugation, carried out with the CST Microwave Studio's finite difference time domain (FDTD) method.  30 GHz performance will be the focus of a future paper once we have complete fabrication of 30 GHz arrays.

\begin{figure}[htbp]
\begin{center}
\includegraphics[width=1\linewidth, keepaspectratio]{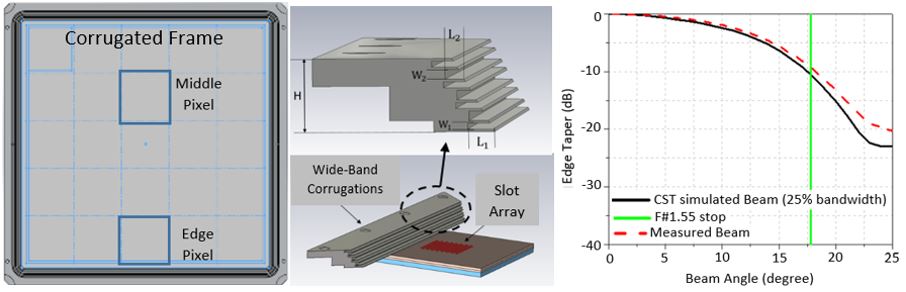}
\caption{{\it Left:} View of the 40 GHz focal plane layout shows 16 edge and 9 middle antenna arrays, wide-band corrugated frame with 3λ/8 distance to the edge antenna array, {\it Center:} Slot-antenna array with all dielectric stacks and the corrugated frame in the CST commercial software for vertical polarization (Pol A), and {\it right:} The expected and measured 1D beam profiles of edge antenna array at 40 GHz. The simulated 1D pattern is calculated over 25\% bandwidth. The simulated and measured half width at full maximum (FWHM) are 10.9 degree and 10.25 degree, respectively. The F\#1.55 stop for BICEP Array subtends 17$^o$. (Color figure online.)}

\end{center}
\label{fig3}
\end{figure}

We have measured far field patterns of 40 GHz antennas in the presence of the wide-band  corrugated  frame, and we test the performance of the corrugations by comparing beams from antennas on the tile perimeter adjacent to the frame to those from the interior of the tiles. Beam maps are taken on the detectors by illuminating them with a chopped (10-25 Hz) thermal source moved through a grid-like scan. We voltage bias our detectors onto an aluminum superconducting transition designed for high optical loading and we demodulate the time stream current data using a chopper reference signal to minimize noise. We show only the in-phase quadrature of the demodulation to avoid noise-biasing the resulting maps.  Fig. 3 shows that the measured beams averaged over the tile (red dashed line) agree well with the simulated beam profile (solid black), and that they match the f/1.55 optics. 

Difference maps are highly sensitive to interactions with the module frame.  Figure 4 left shows the difference between perimeter detectors on a 40 GHz tile with the new corrugation and a BICEP3 tile with the narrow band corrugations.  We spatially rescale the 95 GHz beam to match the size of the 40 GHz beam and normalize both to their peak values.  The $\sim \pm$ 10\% difference between the maps suggests that one beam (95 GHz) is highly steered away from the other.  We test that the steering by frame interactions is much less in the 40 GHz in Fig. 4 right, which compares the interior and exterior beams.  This difference is only $\sim \pm$ 3\%.

We fit a two dimensional elliptical Gaussian profile to the main beam of each detector [7]:    

\begin{equation}
	B(\bm{x}) = \frac{1}{\mathrm{\Omega}}e^{-\frac{1}{2}(\bm{x}-\bm{\mu})^{T}\Sigma^{-1}(\bm{x}-\bm{\mu})},
\end{equation}
where $\mathrm{\Omega}$ is the normalization, and $\bm{x} $ is the beam map coordinate, $\bm{\mu} = (x_0,y_0)$ is the beam center.  Figure 5 shows the Gaussian fits (second and fourth rows) to measured beams of each polarization (first and third rows), as well as their in-pixel differences in the right most column.  The top rows of the chart shows a pixel in the tile's interior, while the bottoms shows an edge pixel.  They have comparable dipolar mismatch, suggesting that the corrugated frame has limited effect on beam synthesis.  The histograms of offsets in Fig. 6 further supports this conclusion.

\begin{figure}[htbp]
\begin{center}
\includegraphics[width=0.9\linewidth, keepaspectratio]{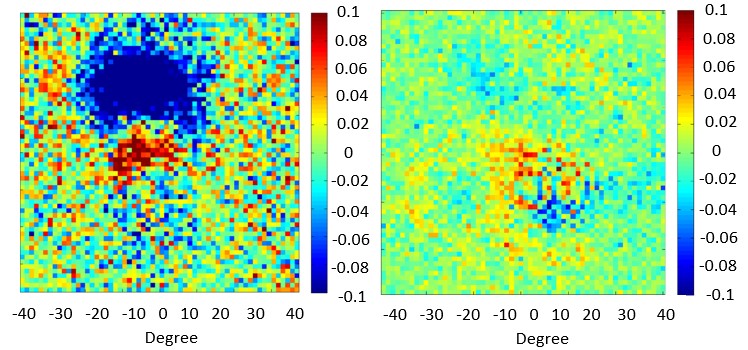}
\caption{ {\it Left:}  Measured difference between a 40 GHz BICEP Array pixel and a spatially rescaled 95 GHz BICEP3 Array pixel where the corrugations were not as carefully optimized.  The difference suggests strong differential pointing of one beam relative to the other.  {\it Right:}  Difference of an edge and interior 40 GHz beam, showing a reduced difference and suggesting that the point in the {\it left} map is from the 95 GHz antenna's frame interactions.  All maps are peak normalized before subtraction. (Color figure online.)}

\end{center}
\label{fig4}
\end{figure}

\begin{figure}[htbp]
\begin{center}
\includegraphics[width=0.8\linewidth, keepaspectratio]{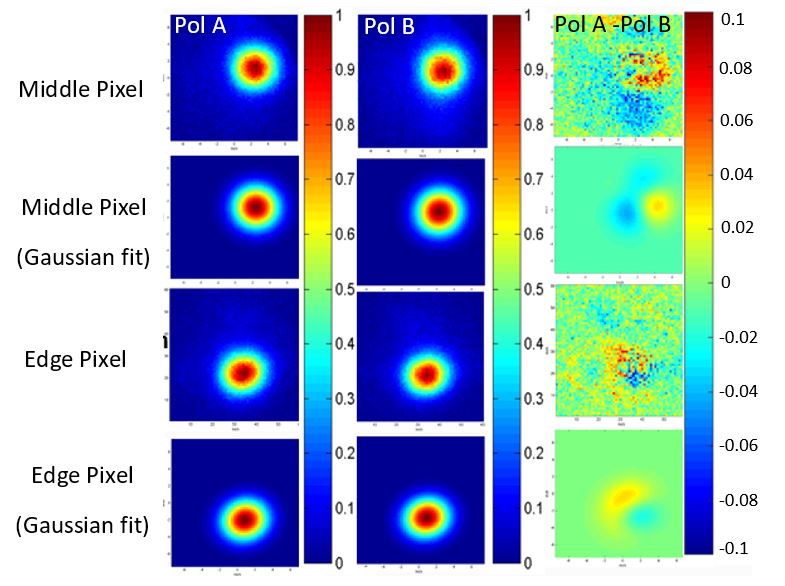}
\caption{ The measured contour plot of Pol A, Pol B and beam map difference for two working middle and edge antenna array pairs over ~ 25\% bandwidth, and the corresponding Gaussian fit.(Color figure online.)}

\end{center}
\label{fig5}
\end{figure}

\begin{figure}[htbp]
\begin{center}
\includegraphics[width=0.8\linewidth, keepaspectratio]{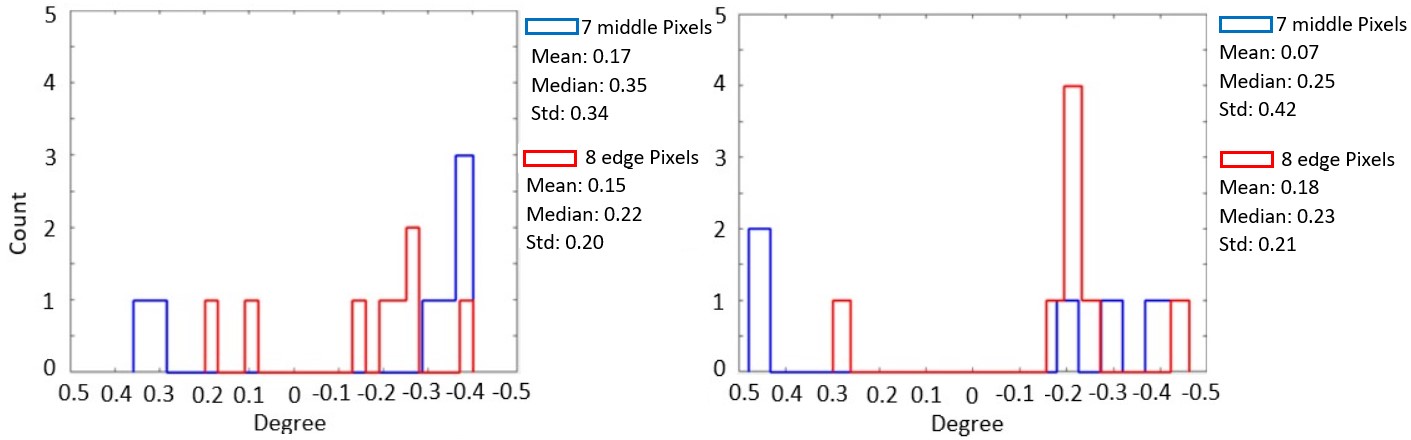}
\caption{ {\it Left:}  histogram of differential pointing of Pol A-Pol B in vertical axis,  {\it Right:} histogram of differential pointing of Pol A-Pol B in horizontal axis for 7 middle and 8 edge antenna array pair. The Gaussian differential pointing values for both pairs are within the same values which indicates a smaller beam mismatch contributions from the corrugated frame. (Color figure online.)}

\end{center}
\label{fig6}
\end{figure}

\section{Optical Efficiency Enhancement}

This section shows the effect of finite array simulation to improve the detector optical efficiency. 
The impedance and efficiency plots in Fig. 2 draw from simulations that model a single radiator surrounded by master/slave boundary conditions and constant input feed impedance matching.  These boundary conditions effectively model an infinitely large array.  In practice, the antenna is finite in size and impedance varies from the center to the edge, most closely matching the infinite case for the center slots. This results in an impedance mismatch losses. We simulated an 8X8 array in the CST commercial software, as shown in Fig. 7. There are significant variations in radiation impedance, but surprisingly stable reactance, consistent with other studies [8].  The software computes an S-matrix between excitations at lumped ports over the slots and a free traveling plane wave above the antenna.  We can also construct an S-matrix from a simple circuit model of the feed network that describes coupling between the slot feeds and the bolometer.  Cascading the two matrices results in an aggregate efficiency between a free space plane wave and the bolometers.  We compute that a feed network designed to match the infinite array impedance but connected to the finite array achieves a respectable 84\% efficiency.  However, by numerically varying the feed point impedance uniformly across the array (i.e. no spatial variations), we improve the optical efficiency to 93\%. We intend to implement this in a future design.

\begin{figure}[htbp]
\begin{center}
\includegraphics[width=0.7\linewidth, keepaspectratio]{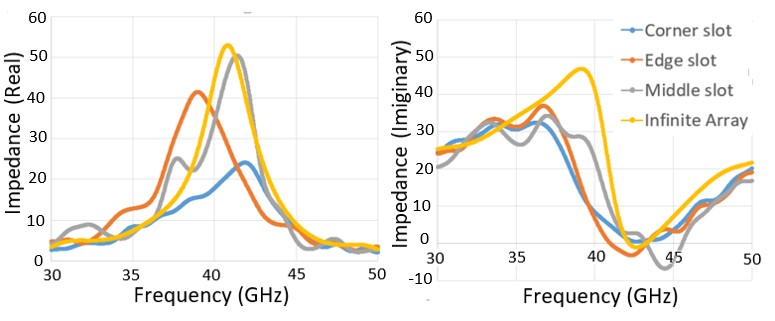}
\caption{ {\it left:} The expected  real impedance for infinite array versus finite array (edge vs middle vs corner slots). the impedance is about factor of 2 lower for the corner slot compared to the middle slot in the array.{\it Right:} The expected imaginary impedance for infinite array versus finite array. (Color figure online.)}

\end{center}
\label{fig7}
\end{figure}

\pagebreak
\section{Conclusion and Future Work} 

The measured optical performance of the 40 GHz detectors is consistent with simulations.  Moreover, our detector development efforts have greatly reduced the near field beam mismatch to below 8\% peak to peak, which is subdominant to our prior studies of systematic errors [7]. The minimized near-field
mismatch might be translated to the far field of the camera [5] due to non-idealities in the optics design system. This will reduce the beam residuals leakage ($T \rightarrow P$) in CMB science maps and bias constraints on $r$ at levels far below BICEP Array's target sensitivity of 0.01.  In this sense, the detector performance is acceptable. We also proposed a new approach to enhance the optical efficiency of the detector to be implemented in our future design. 
We have also developed 30 GHz detector that use a simple rescaling of the antenna pitch and radiator size, but with an impedance matching that optimized for the same 625 $\mu$m silicon wafer thickness.  The 30 GHz detector is being fabricated and we are planning to deploy it to South Pole by end of 2019.  Its performance will be the subject of a future paper.

\begin{acknowledgements}
(The BICEP/Keck project have been made possible through a series of grants from the National Science Foundation including 0742818, 0742592, 1044978, 1110087, 1145172, 1145143, 1145248, 1639040, 1638957, 1638978, 1638970, {\&} 1726917 and by the Keck Foundation. The development of antennacoupled detector technology was supported by the JPL Research and Technology Development Fund and NASA Grants 06-ARPA206-0040, 10-SAT10-0017, 12-SAT12-0031, 14-SAT14-0009 {\&} 16-SAT16-0002. The development and testing of focal planes were supported by the Gordon and Betty Moore Foundation at Caltech. Readout electronics were supported by a Canada Foundation for Innovation grant to UBC. The computations in this paper were run on the Odyssey cluster supported by the FAS Science Division Research Computing Group at Harvard University. The analysis effort at Stanford and SLAC is partially supported by the U.S. DoE Office of Science. We thank the staff of the U.S. Antarctic Program and in particular the South Pole Station without whose help this research would not have been possible. Tireless administrative support was provided by Kathy Deniston, Sheri Stoll, Irene Coyle, Donna Hernandez, and Dana Volponi
\end{acknowledgements}


\begin{thebibliography}{99}

\bibitem{BK15}
Bicep/Keck Collaboration {\it Phys. Rev. Lett.} \textbf{121}, 22, (2018), 
DOI: 10.1103/PhysRevLett.121.221301

\bibitem{Ale19}
A. Schillaci, et al, {\it J. Low Temp. Phys.}, This Special Issue (2019)

\bibitem{Hui18}
H. Hui, et al, {\it Proc. SPIE} \textbf{10708}, id. 1070807, (2018), 
DOI:  10.1117/12.2311725.

\bibitem{BKAntenna}
The \textit{Keck Array} and BICEP2 Collaborations, {\it The Astrophysical Journal,} DOI: 10.1088/0004-637x/812/2/176(2015).

\bibitem{BKXI}
The \textit{Keck Array} and BICEP2 Collaborations, arXiv e-prints (2019), arXiv:1904.01640 

\bibitem{Soliman18}
A. Soliman, et al, {\it Proc. SPIE} \textbf{10708}, id. 107082G, (2018), 
DOI:  10.1117/12.2312942


\bibitem{Roger}
The \textit{Keck Array} and BICEP2 Collaborations, {\it The Astrophysical Journal,}
DOI:10.1088/0004-637X/806/2/206 (2015). 

\bibitem{H11}
H. Motevasselian, et al, {\it Proc. IEEE} \textbf{10708}, Id. 6050307, (2011).

\end{thebibliography}
\end{document}